\documentclass[epj,twocolumn,floatfix,final]{svjour}
\usepackage{graphicx}  
                                               
\begin{document}
   
%\twocolumn[\hsize\textwidth\columnwidth\hsize\csname@twocolumnfalse\endcsname
 
%\title[Jet formation in a collapsing Bose-Einstein condensate]{Mean-field
%model of jet formation in a collapsing Bose-Einstein condensate}

\title{Bright solitons and soliton trains in a fermion-fermion mixture}

\author{Sadhan K. Adhikari\thanks{e-mail: adhikari@ift.unesp.br}}
%\affiliation{Instituto de F\'{\i}sica Te\'orica, Universidade Estadual
%Paulista, 01.405-900 S\~ao Paulo, S\~ao Paulo, Brazil}
%\address{Instituto de F\'{\i}sica Te\'orica, Universidade Estadual
%Paulista,  01.405-900 S\~ao Paulo, S\~ao Paulo, Brazil}
\institute{Instituto de F\'{\i}sica Te\'orica, UNESP $-$ S\~ao Paulo State
University,  01.405-900 S\~ao Paulo, S\~ao Paulo, Brazil}

\date{\today}

\abstract{We use a time-dependent dynamical mean-field-hydrodynamic
model to predict and study  bright solitons in a degenerate
fermion-fermion mixture in a quasi-one-dimensional cigar-shaped geometry
using variational and numerical methods.
Due to a strong Pauli-blocking repulsion among identical spin-polarized
fermions at short distances there cannot be bright  solitons for
repulsive interspecies fermion-fermion interactions.  However, 
stable bright solitons can be formed for a sufficiently
attractive interspecies  interaction. We perform a 
numerical stability analysis of these solitons and also demonstrate the
formation of soliton trains.  These fermionic solitons can be formed and
studied in  laboratory with present technology.}

%\PACS{{45.05.+x}{General theory of classical mechanics of discrete
%systems} \and {05.45.-a}{Nonlinear dynamics and chaos} \and
%{03.75.Hh}{Static properties of condensates; thermodynamical,
%statistical, and structural properties}}

%\maketitle
 

 \authorrunning{S. K. Adhikari}
\titlerunning{Bright solitons and soliton trains in a fermion-fermion
mixture}
\maketitle

Recent observations  \cite{exp1,exp2,exp3,exp4}  and associated
experimental \cite{exp5,exp5x,exp6} and theoretical
\cite{yyy1,yyy,capu,capu1,ska} studies
of  a degenerate Fermi gas (DFG) by
sympathetic cooling in
the presence of a second boson or fermion component suggest the
possibility of soliton formation. 
Apart from the observation of a DFG in the following  
degenerate boson-fermion
mixtures (DBFM)
 $^{6,7}$Li
\cite{exp3}, $^{23}$Na-$^6$Li \cite{exp4} and $^{87}$Rb-$^{40}$K
\cite{exp5,exp5x}, there have been studies of degenerate 
spin-polarized fermion-fermion
mixtures (DFFM)
 $^{40}$K-$^{40}$K \cite{exp1} and $^6$Li-$^6$Li \cite{exp2}.

Bright solitons in a Bose-Einstein condensate
(BEC) are formed due to an attractive nonlinear atomic
interaction \cite{exdks}.  As the interaction in a pure DFG at short
distances is
repulsive due to strong Pauli blocking, there cannot be bright solitons in
a DFG. 
However, it has been demonstrated \cite{fbs1,fbs2} that  bright solitons
can be
formed in a DBFM in the presence of a sufficiently strong
boson-fermion attraction which can overcome the Pauli repulsion among
identical fermions.

We demonstrate the formation of stable fermionic bright solitons in a DFFM
for a sufficiently attractive interspecies fermion-fermion interaction.  
In a DFFM, the coupled system can lower its energy by forming high density
regions, the bright solitons, when the attraction between the two types of
fermions is large enough to overcome the Pauli repulsion among identical
fermions. We use a coupled time-dependent mean-field-hydrodynamic model
for a DFFM and consider the formation of axially-free localized bright
solitons in a quasi-one-dimensional cigar-shaped geometry using numerical
and variational solutions.  The present model is inspired by the success
of a similar model suggested recently by the present author in the
investigation of collapse \cite{ska} and bright \cite{fbs2} and dark
\cite{fds} solitons in a DBFM. We study the condition of modulational
instability of a constant-amplitude solution in this model and demonstrate
the possibility of the formation of bright solitons. We also present a
numerical stability analysis of these robust bright solitons and consider
the formation of a soliton train in a DFFM by a large sudden jump in the
interspecies fermion-fermion scattering length near a Feshbach resonance,
experimentally observed in both $^6$Li-$^6$Li and $^{40}$K-$^{40}$K
\cite{fesh}.

We  use a  simplified mean-field-hydrodynamic Lagran\-gian for 
a DFG used successfully to study a DBFM
\cite{ska,fbs2,fds}. 
The virtue of the
mean-field model over a  microscopic description is its simplicity and
predictive power. 
To  develop a set of  time-dependent
mean-field-hydrodynamic
equations for the interacting DFFM, we use   the
following Lagrangian density \cite{ska,fbs2} 
\begin{eqnarray}\label{yy} &{\cal
L}&= g_{12}n_1n_2+\sum_{j=1}^2
\frac{i}{2}\hbar \left[ \psi_j\frac{\partial {\psi_j} ^*}{\partial
t} - {\psi_j}^* \frac{\partial \psi_j}{\partial t} \right]
\nonumber \\ 
&+& \sum_{j=1}^2     
\left(\frac{\hbar^2 |\nabla_{\bf r} \psi_j|^2 }{6m_j}+
V_j({\bf r})n_j+\frac{3}{5} A_j n_j^{5/3}\right),
\end{eqnarray} 
where $j=1,2$ represents the two components, $\psi_j$ the 
complex probability 
amplitude, $n_j=|\psi_j| ^2$ the real probability 
density,  
$^*$ denotes complex conjugate,  $m_j$  the
mass,   
$A_j=\hbar^2(6\pi^2)^{2/3}\- \- /(2m_i),$ 
the interspecies coupling        
$g_{12}=2\pi \hbar^2 a_{12} 
/m_R$ 
with $m_R=m_1m_2/(m_1+m_2)$ the reduced mass,  and 
$ a_{12}$ 
 the interspecies 
fermion-fermion scattering length.
 The
number of fermionic atoms $N_j$
is given by  $\int d{\bf r} n_j({\bf r})=N_j$.
The trap potential with axial symmetry is  $
V_{j}({\bf
r})=\frac{1}{2}m_j \omega ^2 (\rho^2+\nu^2 z^2)$ where
 $\omega$ and $\nu \omega$ are the angular frequencies in the radial
($\rho$) and axial ($z$) directions with $\nu$ the anisotropy.
The interaction between identical intra-species fermions in
spin-polarized state is highly suppressed due 
to Pauli blocking terms $3A_jn_j^{5/3}/5$ 
and has been neglected in Eq. (\ref{yy}).   The
kinetic energy terms $\hbar^2|\nabla_{\bf r}\psi_j|^2/(6m_j)$
in Eq. (\ref{yy})
contribute little to this problem compared to the
dominating Pauli-blocking terms.  
However, its inclusion leads
to an analytic solution for the probability density everywhere
\cite{fbs2}.

With the Lagrangian density (\ref{yy}), the following Euler-Lagrange
equations   can be derived in a
straight-forward fashion 
\cite{ska,fbs2}:  \begin{eqnarray}\label{e} \biggr[ 
i\hbar\frac{\partial }{\partial t} +\frac{\hbar^2\nabla_{\bf
r}^2}{6m_{{j}}} - V_{{j}} - A_jn_j^{2/3}-
g_{{12}}
n_k
 \biggr]\psi_j=0,
\end{eqnarray}
where $j\ne k = 1,2$. This is essentially a time-dependent version of a
similar time-independent model used recently for fermions \cite{capu}.
For large $n_j$, both lead to the
Thomas-Fermi result $n_j=[(\mu_j-V_j)/A_j]^{3/2}$ \cite{capu,ska}
with $\mu_j$ the chemical potential.   
As the bright solitons of this 
rapid note are  stationary states, they could be obtained by 
the time-independent approach  used in Ref. \cite{capu}.  The stationary
approach of Ref. \cite{capu}  has passed
rigorous tests of comparison of the hydrodynamic-mean-field  
spectra of localized fermions with the spectra calculated in the
collisionless regime within the random-phase approximation (RPA). The
results of
mixing-demixing and collapse of the hydrodynamic approach are in agreement
with the RPA analysis \cite{capu1}. The detailed behavior of collective
excitation  of trapped fermions has also been found to agree with that
obtained by
an RPA analysis \cite{capu}. For a description of stationary
solitons (e.g., of
Fig. 1) we could have used  the well-established formulation of
Ref. \cite{capu} to obtain identical results, as 
the present time-dependent dynamical description 
and the time-independent approach of \cite{capu} yield  
identical  results
for   stationary states.  However, we shall be  using the present 
time-dependent dynamical formulation to study the nonequilibrium
generation soliton
trains, in addition.

We reduce  three-dimensional Eqs. (\ref{e})  to a minimal 
quasi-one-dimensional form 
in a  cigar-shaped  geometry
with $\nu << 1$, where 
the radial motion is frozen in the ground state of the harmonic trap and
the dynamics is carried by the axial motion. 
For radially-bound and axially-free solitons we eventually set 
$\nu =0$.
Following Ref. \cite{fbs2} this reduction can be done
in a
straight-forward fashion and we quote the final results here: 
\begin{eqnarray}\label{m} \biggr[ i\frac{\partial
}{\partial \tau}
+\frac{\partial ^2}{\partial y^2} 
-    
 N_{jj}
\left|{{\phi}_j}\right|^{4/3}        
+N_{jk}
  \left|{{\phi}_k}\right|^2                  
 \biggr]{\phi}_{{j}}({y},\tau)=0,         
\end{eqnarray}
where $\phi_j, j\ne k=1,2$ represents the two solitons,  $\tau =t
\omega/2$, $y=z/l$, 
$N_{jj}=9(6\pi N_j)^{2/3}/5, $
$N_{jk}=12|a_{12}|N_k/l,$    $l=\sqrt{\hbar/(\omega m)}$, with
$m=3m_1=3m_2$. Here we employ  
equal-mass fermions,
a negative $a_{12}$ corresponding to
attraction, and  normalization   
$\int_{-\infty}^\infty |\phi_j(y,\tau)|^2 dy =1 .$
In Eqs. (\ref{m})  
a sufficiently strong  
attractive  fermion-fermion coupling
 $N_{jk}|\phi_k|^2 (j\ne k)$ can  overcome  the Pauli 
repulsion $N_{jj}|\phi_j|^{4/3}$
and  form  bright solitons.    

\begin{figure}%[!ht]
 
\begin{center}
\includegraphics[width=.9\linewidth]{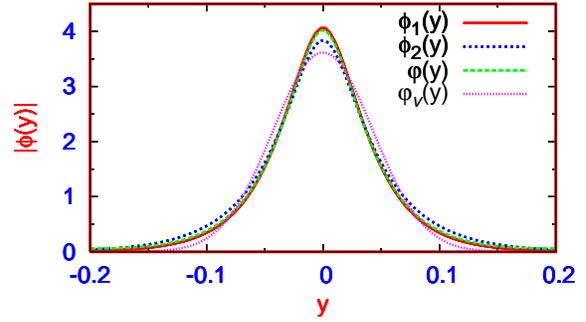}
\end{center}

\caption{(Color online)  The 
solitons  $|\phi_j(y)|$  of Eq. (\ref{m})
  vs. $y$ (in dimensionless units)
for 
 $N_1=44$, $N_2=56$, $a_{12}=-0.3 $ nm, while   
$N_{11} \approx 159 $, $N_{12}\approx  203$, $N_{21}\approx  160$, and
$N_{22}\approx 187$. 
The variational $(\varphi_v)$ and  numerical 
$(\varphi)$ solutions of
Eq. (\ref{o}) 
for $N_1=N_2=50$  are also shown. 
} \end{figure}

\begin{figure}%[!ht]
 
\begin{center}
\includegraphics[width=.8\linewidth]{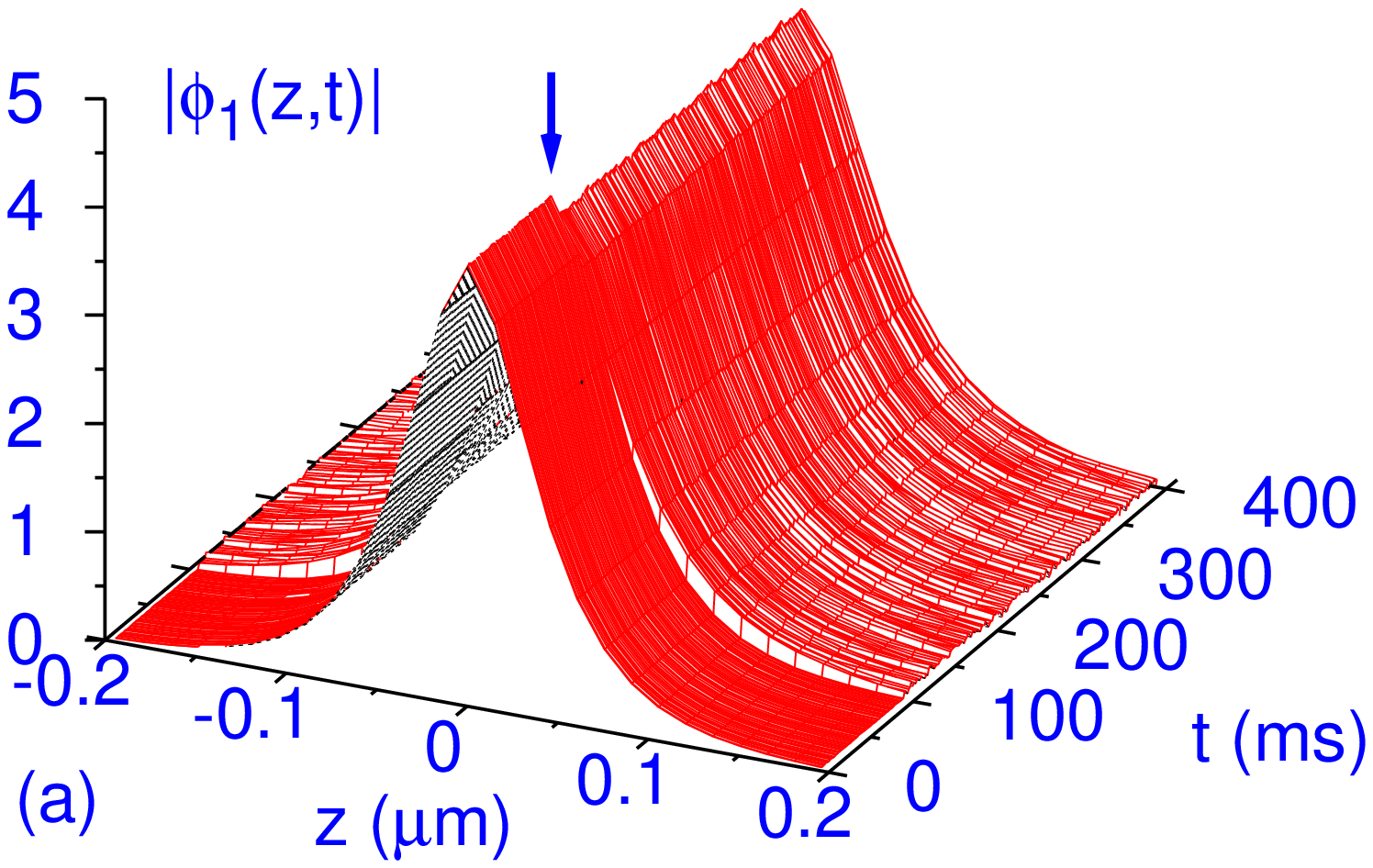}
\includegraphics[width=.8\linewidth]{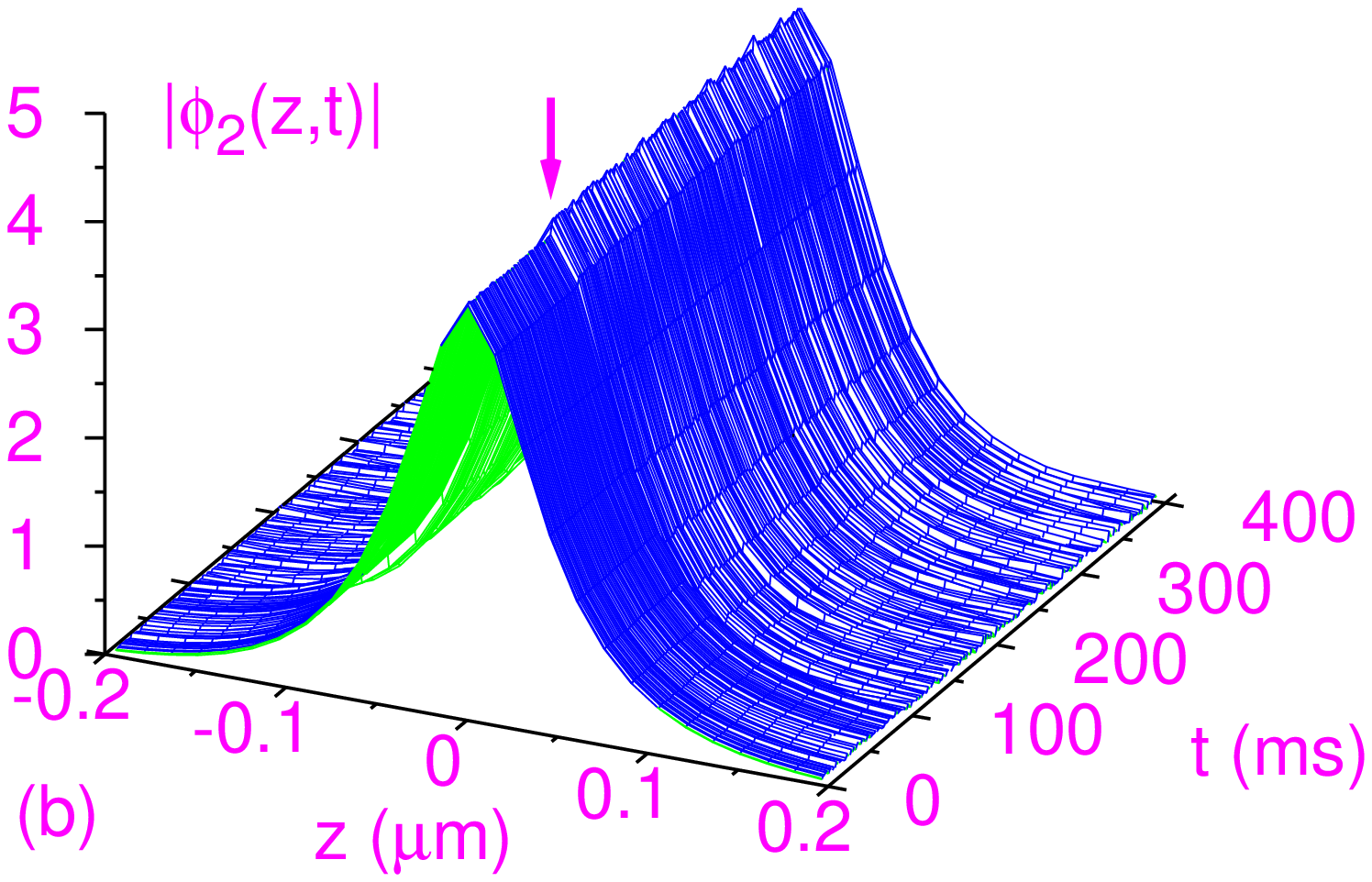}
\end{center}

\caption{(Color online)  The propagation of fermionic
solitons (a)  $|\phi_1(z,t)|$ and (b)  $|\phi_2(z,t)|$ of Fig. 1 
  vs. $z$  and $t$.
At
$t=100$ ms (marked by arrows) the bright solitons are set into small
breathing oscillation by suddenly  changing  $N_1=44$ and  $N_2=56$
to  $N_2=44$, $N_1=56$.} \end{figure}

Now we perform a
stability analysis  of constant-ampli\-tude 
solutions of Eqs. (\ref{m}) 
 and  study the
possibility of generation 
of  solitons 
in the symmetric case: $N_1=N_2$, when 
$\phi_1=\phi_2 \equiv \varphi$ and these equations reduce to
\begin{eqnarray}\label{o} \biggr[ i\frac{\partial
}{\partial \tau}
+ \frac{\partial ^2}{\partial y^2} 
-    \beta
\left|{\varphi}\right|^{4/3}        
+\gamma
  \left|{{\varphi}}\right|^2                  
 \biggr]{\varphi}({y},\tau)=0,         
\end{eqnarray}
where $\beta=N_{11}=N_{22}$ and $\gamma=N_{12}=N_{21}$. 
We consider the 
constant-amplitude solution \cite{1}
$\varphi_0=A_0
\exp(i\delta) \equiv A_0 \exp [i(\gamma A_0^2 \tau -\beta A_0^{4/3}
\tau)]$ of Eq. (\ref{o})
under small perturbation: $\varphi=(A_0+ A)\exp(i \delta)$,
where $A=A(y, \tau)$ and  $A_0$ the amplitude. Substituting this
perturbed solution in Eq. (\ref{o}), and for small perturbations retaining
only the linear terms in $A$ we get 
\begin{eqnarray}\label{p} i\frac{\partial A
}{\partial \tau}
+ \frac{\partial ^2 A}{\partial y^2} 
-   \frac{2}{3} \beta A_0^{4/3}(A+A^*)
+\gamma A_0^2 (A+A^*)=0.
\end{eqnarray}
We consider the plane-wave
perturbation  $A(y,\tau)=
A_1 \- \cos (K\tau -\Omega y)+i A_2 \sin  (K\tau -\Omega y)$ in Eq. 
(\ref{p}). Then separating the real and imaginary terms and 
eliminating  $A_1$ and $A_2$ we obtain the dispersion
relation 
$K=\pm  \Omega [\Omega^2$ $-(2\gamma A_0^2 -4\beta
A_0^{4/3}/3)]^{1/2}.$
For stability of the plane-wave perturbation,  
$K$ has to be real.  This happens for 
$2\gamma A_0^2 < 4\beta A_0^{4/3}/3$ or $\gamma A_0^{2/3}
< 2\beta /3$.  However, $K$ can become imaginary for
$\gamma A_0^{2/3}
> 2\beta /3$  and the plane-wave perturbations can grow exponentially
with time $\tau$. This is the domain of modulational instability 
of a constant-intensity solution, signalling a tendency of 
spatially localized bright
solitons to appear. We also performed this analysis 
 in the case of
non-symmetric coupled equations (\ref{m})  and quote the
result here. The
condition for instability is
$N_{12}N_{21}A_{10}^{2/3}A_{20}^{2/3}> 4 N_{11}N_{22}/9$ \cite{shuk},
where $A_{10}$
and $A_{20}$ are the amplitudes of the two solutions.

Next we present a 
variational analysis of Eq. (\ref{o}) based on the
normalized Gaussian trial wave function \cite{and}
$\varphi_v(y,\tau)$ $ =\sqrt{1/[a(\tau)\sqrt
\pi]}\exp[-y^2/\{2a^2(\tau)\}+i
b(\tau)y^2/2]$, where $a$ is the width and   $b$ the chirp. 
The Lagrangian density for Eq. (\ref{o})
is the one-term version of Eq. 
(\ref{yy}), which is evaluated with this 
trial function and the effective
Lagrangian  $L
=\int_{-\infty}^\infty {\cal L}(\varphi_v)dy$ becomes 
\begin{equation}
L =\frac{a^2}{4}\biggr(
\dot b+\frac{2}{a^4}+2b^2 -\frac{\sqrt 2}{\sqrt \pi} \frac{\gamma
}{a^3}+ \frac{12 \sqrt
3}
{5\sqrt 5}\frac{\beta  }{\pi ^{1/3}a^{8/3}} \biggr).
\end{equation}
The  variational  Euler-Lagrangian
equations
for $a$ and  $b$ 
can then be written and solved 
in a standard fashion \cite{and} to
yield the differential equation for the width: 
$d^2 a/d\tau^2 = 
[4- a\gamma \sqrt{2/\pi}+a^{4/3}(8\beta \sqrt 3)/(5\pi^{1/3}
\sqrt5)]/a^3$. The variational result for width $a$ follows by setting the
right hand side of this equation to zero, from which the variational 
profile for the soliton can be obtained \cite{and}.

We solve Eqs.  
 (\ref{m})  for bright  solitons
numerically using a time-iteration
method based on the Crank-Nicholson discretization scheme
elaborated in Ref. \cite{sk1}  
using time step $0.0002$ and space step $0.015$.
We perform  a time evolution of Eqs.  (\ref{m}) 
 introducing an harmonic oscillator potential $y^2$ 
and setting the nonlinear terms to
zero, and  
starting with the eigenfunction of the linear harmonic
oscillator problem.
%: $\phi_i(y,\tau) 
%=
%\pi^{-1/4}\exp(-y^2/2)\exp(-i\tau).$  
The extra 
harmonic oscillator potential, which will be set equal to zero in the end,
only aids in starting the time evolution
with an exact analytic form. 
During the time evolution the nonlinear
terms are  switched on  and  the harmonic oscillator potential is
switched off slowly and 
the time evolution  continued to obtain the final converged
solutions.

\begin{figure}%[!ht]
 
\begin{center}
\includegraphics[width=.8\linewidth]{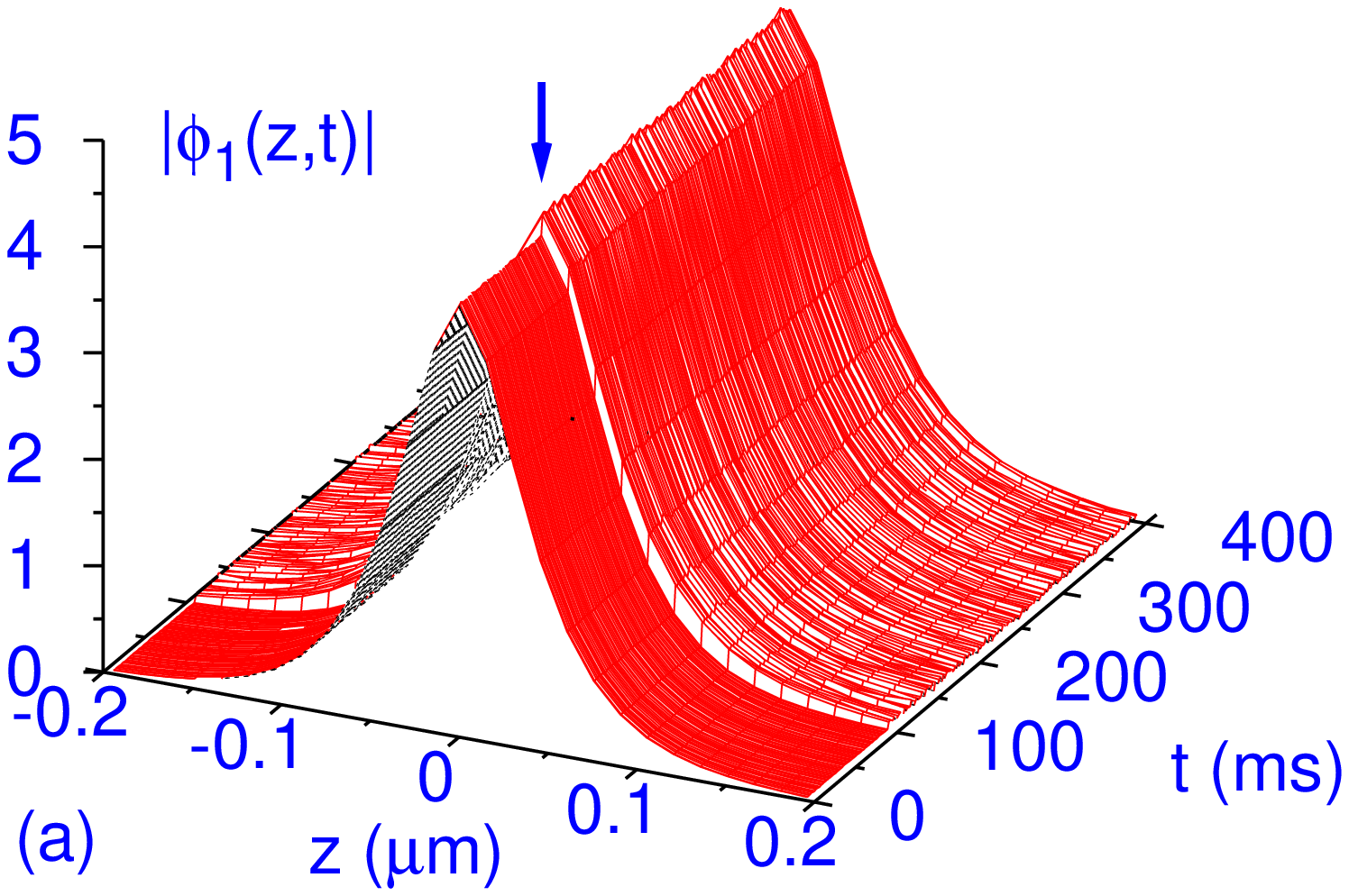}
\includegraphics[width=.8\linewidth]{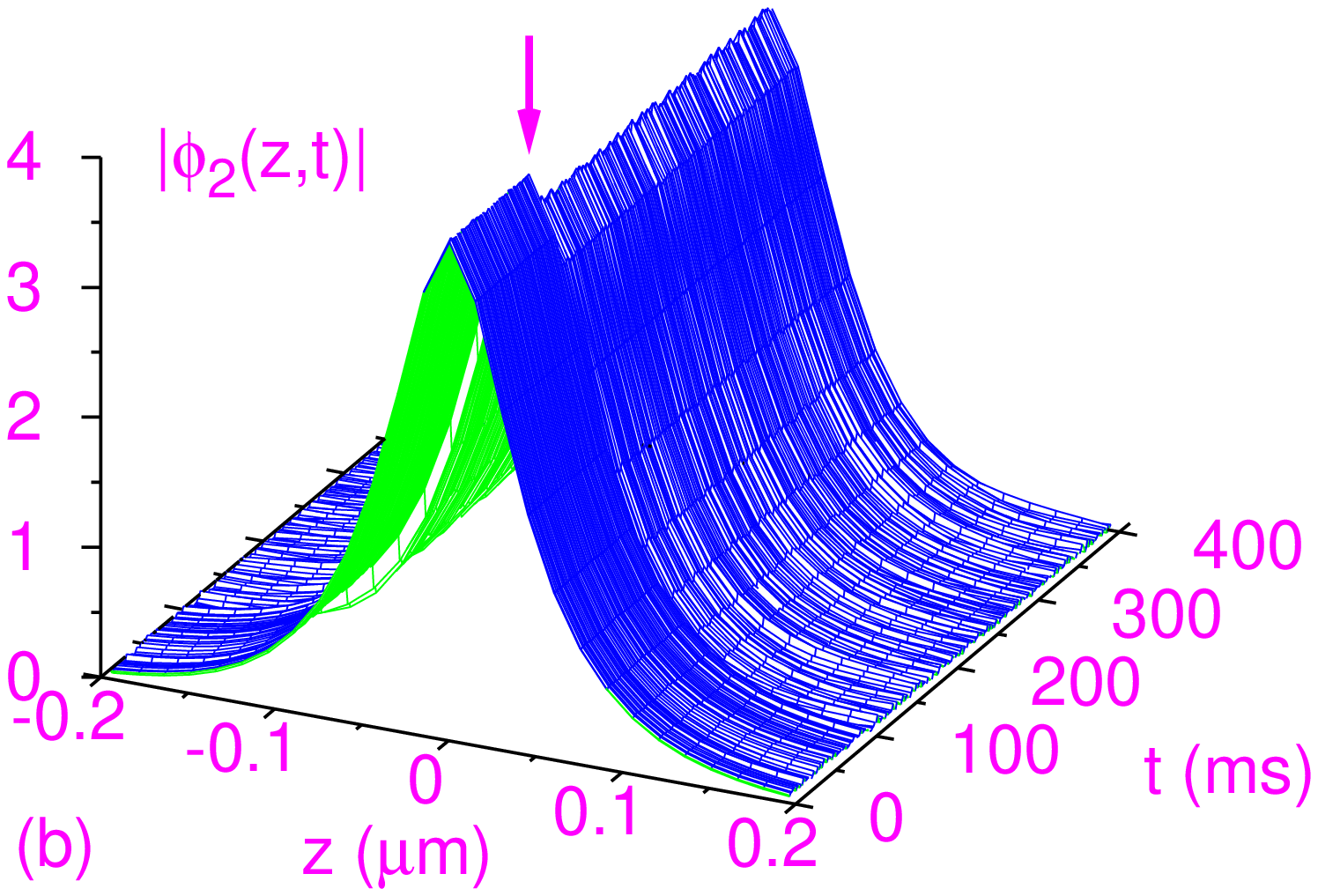}
\end{center}

\caption{(Color online)  The propagation of fermionic
solitons (a)  $|\phi_1(z,t)|$ and (b)  $|\phi_2(z,t)|$ 
  vs. $z$  and $t$. At $t=100$ ms (marked by arrows)  the same initial
solitons of Figs. 1
are set into small
breathing oscillation by suddenly  changing  $\phi_1 \to 1.1 \phi_1$ and 
$\phi_2 \to 0.9 \phi_2$.   } \end{figure}

In our numerical study we take $l=1$ $\mu$m and 
consider a DFFM consisting of two electronic states of 
 $^{40}$K  atoms. This  corresponds to a radial
trap of frequency $\omega \approx 2\pi \times 83$ Hz. 
Consequently, the  unit of
time is  $2/\omega \approx 4$ ms.

First we solve coupled Eqs. (\ref{m}) with  
$N_1=44  , N_2= 56$, and $a_{12}=-0.3$ nm.
The soliton profile in this case is shown in Fig. 1, where we also plot
the variational and numerical solutions of Eq. (\ref{o}) 
for $N_1=N_2=50$. In this case in  Eq. (\ref{o})    
$\beta \approx 173$ and   $\gamma \approx 181.6$
leading to a variational
width $a \approx 0.043$ and a   variational soliton profile 
$\varphi_v \approx 3.62\exp(-270y^2)$. The variational result agrees 
well with the numerical solutions.

At this stage it is pertinent to see if Friedel oscillations \cite{fri}
of density of the localized fermions are small so that the effective
description of the solitons is valid. An one-dimensional degenerate Fermi
gas of $N$ atoms filled up to Fermi sea has a spatial extension $2L_F= l
\sqrt{2N-1},$ where $l$ is a measure of confinement \cite{gle}. In the
presence of an harmonic trap, $l$ is the harmonic trap length and is
smaller than the spatial extension of the confined fermions.  In the case
of the soliton of Fig. 2, a typical measure of $l$ could be 0.05 $\mu$m so
that for about 50 fermions considered here  $2L_F = 0.5$ $\mu$m. The Fermi
momentum of an one-dimensional Fermi gas is $k_F=\pi N/(2L_F)$ \cite{gle}. 
The spatial
wave-length of Friedel oscillation is \cite{gle} $\lambda=\pi/k_F=2L_F/N
\approx
0.01$ $\mu$m, much smaller than the soliton width of 0.1 $\mu$m. This
qualitative analysis resulting in small   Friedel oscillation 
supports the effective description used  this rapid note for a DFFM. 

To test the robustness of these solitons we
inflicted different perturbations on  them and studied the resultant
dynamics numerically. First, after the formation of the solitons we
suddenly changed the fermion numbers from $N_1=44$  and  $N_2= 56$ to
$N_1=56, N_2= 44$ 
at time $t= 100$ ms.  This corresponds to a sudden
change
of nonlinearities from $N_{11}\approx 159 $, $N_{12}\approx  
203$, $N_{21}\approx 
160$, and $N_{22} \approx 187$ to 
$N_{11}\approx 187 $, $N_{12}\approx  160$, $N_{21}\approx 203$, and
$N_{22} \approx 159$.
The resultant dynamics is 
shown in  Figs. 2 (a) and (b). 
Due to
the sudden change in nonlinearities the fermionic bright
solitons are set into stable non-periodic small-amplitude
breathing oscillation.   
Next on the same initial solitons of Figs. 2,  at $t=100$ ms, we suddenly
inflict the perturbation $\phi_1 \to 1.1 \phi_1$ and  $\phi_2 \to 0.9
\phi_2$ and follow numerically the time evolution. 
The result of simulation is shown in Figs. 3 (a) and (b). 
We find that the
solitons again continue  non-periodic breathing oscillation and stabilize
at large times.  
We also gave a small displacement between the
centers of these solitons. 
We
find that after oscillation and dissipation the solitons again come back
to the stable configuration.

\begin{figure}%[!ht]
 
\begin{center}
\includegraphics[width=.8\linewidth]{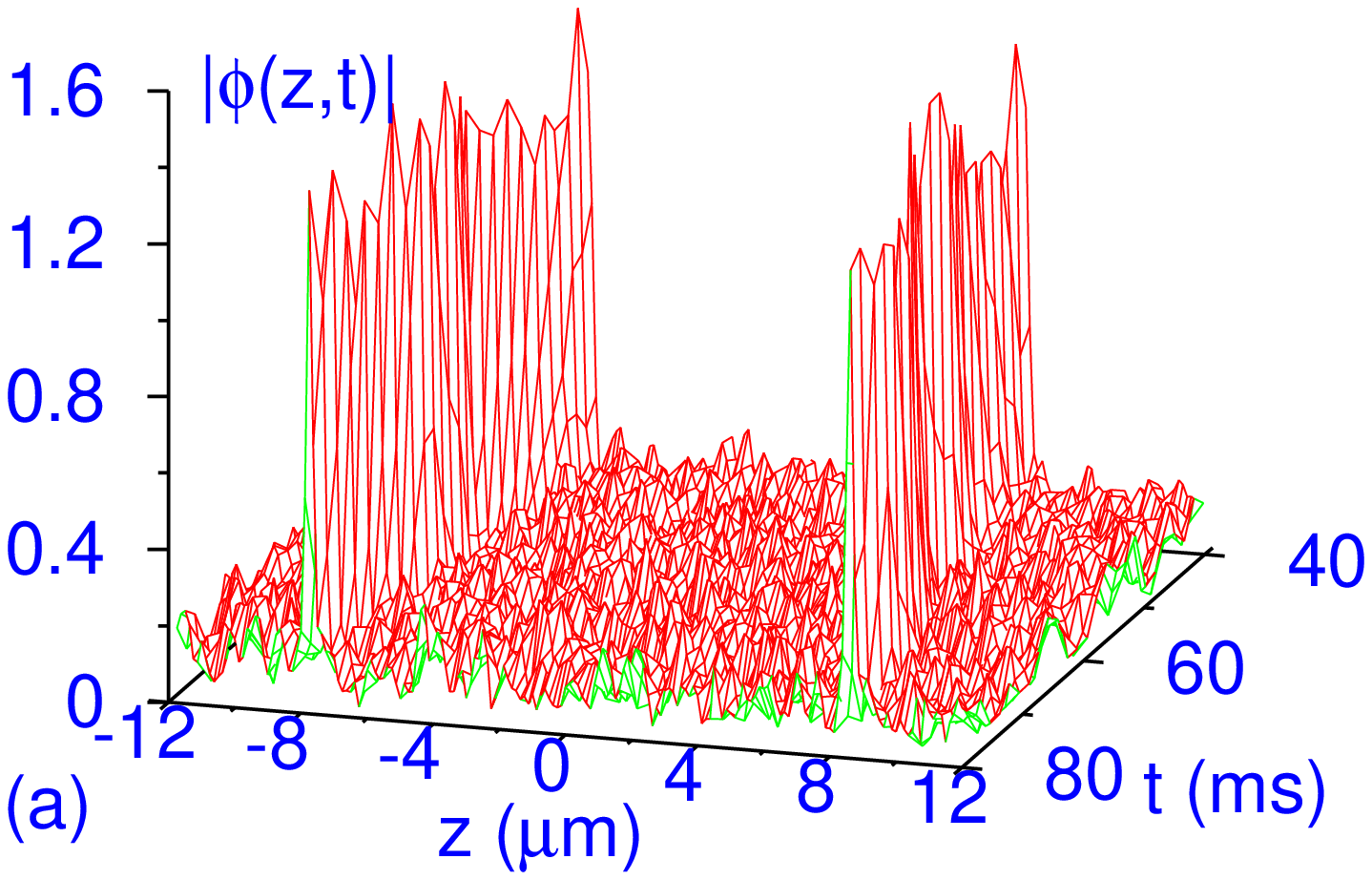}
\includegraphics[width=.8\linewidth]{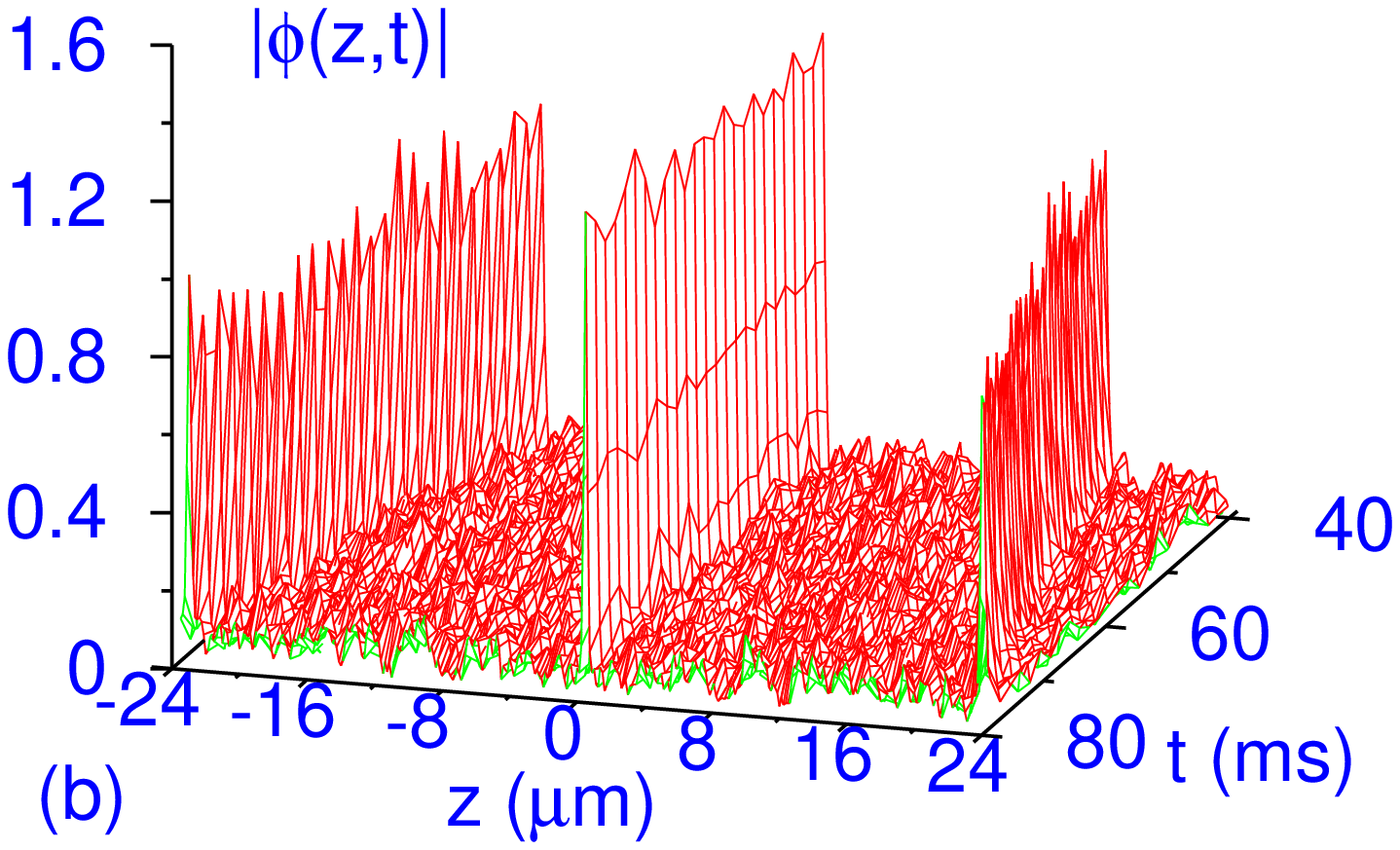}
\end{center}

\caption{(Color online) Soliton trains of  two  and  three solitons
formed upon removing the harmonic trap $y^2$ and 
jumping the nonlinearities at 
$t=0$ from $N_{jj}\approx 173, N_{jk}\approx 145,
k\ne j=1,2$ to (a)  $N_{jj}\approx 173, N_{jk}\approx 273,
$ and to (b) $N_{jj}\approx 173, N_{jk} \approx 309,
$, respectively. } \end{figure}

During the time evolution of Eqs. (\ref{m}) 
if the nonlinearities  are changed 
by a small amount or changed slowly, usually one gets a single stable
soliton when the final nonlinearities are appropriate. 
However,
if the nonlinearities are jumped suddenly by a large amount, a
soliton train can be obtained as in the experiment with BEC \cite{exdks}. 
To illustrate this we consider the solution of 
Eqs. (\ref{m}) with nonlinearities 
$N_{jj} \approx 173$ and $N_{jk}\approx 145, 
j\ne k=1,2$
and harmonic oscillator trap $y^2$. After the formation of the solitons we
suddenly jump the off-diagonal nonlinearities 
to  
$N_{jk} \approx 273$ and also switch off the harmonic trap 
at time
$t=0$. Then after
some initial noise and dissipation 
the
time evolution of  Eqs. (\ref{m}) generates two slowly receding  
bright 
solitons of each component 
as shown in Fig. 4 (a). 
More solitons can be generated when the jump in the nonlinearities is
larger. In Fig. 4 (b) we show the generation of three slowly receding
solitons  of each component     
upon a
sudden jump of the off-diagonal nonlinearities to
$N_{jk} \approx 309$ from the same initial state as in
Fig. 4 (a). 
The formation of soliton trains from a stable initial state is due to
modulational instability \cite{1}.
The sudden jump in the off-diagonal nonlinearities could be
effected by a jump in the interspecies scattering length $a_{12}$ obtained
by
manipulating a background magnetic field near a 
fermion-fermion Feshbach resonance \cite{fesh}.

In conclusion, we use a coupled  
mean-field-hydrodyna\-mic model
for a DFFM to study the formation of bright
solitons and soliton trains in a quasi-one-dimensional geometry
by numerical and variational methods. We find
that an attractive 
interspecies 
interaction can overcome the Pauli 
blocking repulsion and form 
fermionic bright solitons in a DFFM.
The stability of the present solitons is
demonstrated numerically through
their sustained breathing oscillation  
initiated by a sudden small
perturbation. We also illustrate the creation of soliton trains upon
a sudden large jump in off-diagonal
nonlinearities.  Bright solitons and soliton trains
have
been created
experimentally in attractive
BECs in the presence of a  radial trap only without
any
axial trap \cite{exdks}. 
In view of this, fermionic 
bright solitons  and trains can be created in
laboratory in a DFFM in a quasi-one-dimensional configuration. 
Here
we used a set of mean-field equations for 
the DFFM. 
A proper treatment of a DFG or DFFM should be done using a 
fully
antisymmetrized many-body Slater determinant wave 
function
\cite{yyy1,fbs1} as in the case of 
scattering involving many electrons \cite{ps}. However, in
view of the success of a fermionic  mean-field-hydrodynamic  model in
studies of collapse \cite{ska}, bright \cite{fbs2} and dark solitons
\cite{fds} 
in a DBFM, and of
mixing-demixing \cite{md} and black solitons \cite{lpl}  in a DFFM,
we do
not believe that the present study on bright solitons in a DFFM  to be so
peculiar as to have no general validity.

%\acknowledgments

The work is 
supported in part by the CNPq and FAPESP
of Brazil.

%\section*{References}

\end{document}